\documentclass{natureprintstyle_davide}

\usepackage{amsmath,amssymb}
\usepackage{mathpazo} 
\usepackage{bm}
\usepackage{graphicx,subfigure,color}
\usepackage{fancyhdr}
\usepackage{pdfpages}
\usepackage{verbatim}
\definecolor{linkcolor}{rgb}{0.0,0.3,0.5}
\usepackage[unicode,colorlinks=true, linkcolor=linkcolor, citecolor=linkcolor, filecolor=linkcolor,urlcolor=linkcolor]{hyperref}
\usepackage[all]{hypcap} 
\usepackage[all]{nowidow}
\usepackage{titlesec}
\titleformat{\subsection}
  {\normalsize\bfseries}{\thesubsection}{1em}{}	
\usepackage{titlesec}
\titleformat{\section}
  {\large\bfseries}{\thesection}{1em}{}

\fancypagestyle{plain}{
\fancyhf{}
\fancyfoot[R]
\thepage

}
\newcommand\orcid[1]{\href{https://orcid.org/#1}{$\!\!$\includegraphics[scale=0.006]{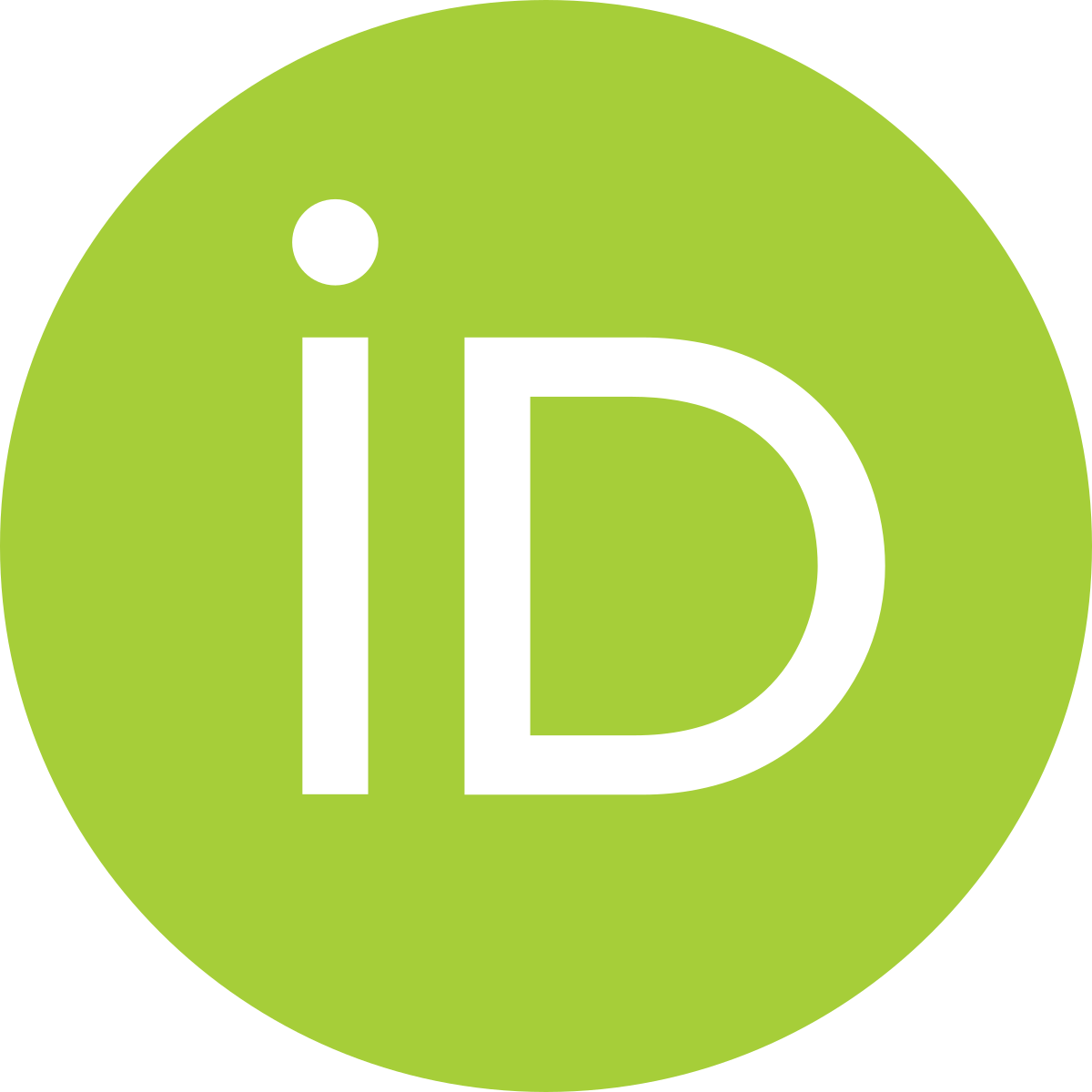} $\!\!$}}
\definecolor{Mygrey}{gray}{0.75}
\newcommand{\ssim}{\mathchar"5218\relax\,} 
\newcommand{\etal}{{\it et al.}}

\usepackage{journals}

\title{Hierarchical mergers of stellar-mass black holes \\ and their gravitational-wave signatures}
\author{\bf 
{\Large Davide Gerosa$\;\;$\orcid{0000-0002-0933-3579}$\,^{1,*}$, 
Maya Fishbach$\;\;$\orcid{0000-0002-1980-5293}$\,^{2}$}
\vspace{0.5cm}
\\
We review theoretical findings, astrophysical modeling, and current gravitational-wave evidence of hierarchical stellar-mass black-hole mergers. While most of the compact binary mergers detected by LIGO and Virgo are expected to consist of first-generation black holes formed from the collapse of stars, others might instead be of second (or higher) generation, containing the remnants of previous black-hole mergers. Such a subpopulation of hierarchically assembled black holes presents distinctive gravitational-wave signatures, namely higher masses, possibly within the pair-instability mass gap, and dimensionless spins clustered at the characteristic value of $\ssim$0.7. In order to produce hierarchical mergers, astrophysical environments need to overcome the relativistic recoils imparted to black-hole merger remnants, a condition which prefers hosts with escape speeds $\gtrsim$ 100 km/s. Promising locations for efficient production of hierarchical mergers include nuclear star clusters and accretion disks surrounding active galactic nuclei, though environments that are less efficient at retaining merger products such as globular clusters may still contribute significantly to the detectable population of repeated mergers.
 While GW190521 is the single most promising hierarchical-merger candidate to date, constraints coming from large population analyses are becoming increasingly more powerful.
\vspace{0.5cm}
}

\begin{document}

\maketitle

\let\thefootnote\relax\footnote{
\begin{affiliations}\rm
\vspace{-0.2cm}

\item School of Physics and Astronomy \& Institute for Gravitational Wave Astronomy, University of Birmingham, Birmingham, B15 2TT, UK.

\item Center for Interdisciplinary Exploration and Research in Astrophysics (CIERA) \& Department of Physics and Astronomy, Northwestern University, 1800 Sherman Ave, Evanston, IL 60201, USA.

 $*$ \href{mailto:d.gerosa@bham.ac.uk}{d.gerosa@bham.ac.uk}
\end{affiliations}
}

\vspace{-1cm}
\section{Introduction}

Gravitational-wave (GW) observations are revolutionizing the field of astronomy and our understanding of compact objects. The prototypical GW sources are merging binaries composed of black holes (BHs) and neutron stars (NSs). At the time of writing, more than 50 of these events have been detected\cite{2016PhRvX...6d1015A,2019PhRvX...9c1040A,2020arXiv201014527A,2019PhRvD.100b3011V,2020PhRvD.101h3030V,2019ApJ...872..195N,2020ApJ...891..123N} by the Advanced LIGO\cite{2015CQGra..32g4001L} and Virgo\cite{2015CQGra..32b4001A} detector network. Exploring the Universe using GWs relies on a deep interplay between astrophysics and relativity. Here we review one of the topics where the dialogue between these two disciplines is most evident: the hierarchical assembly of multiple generations of stellar-mass BH mergers. %

Altough relativistic dissipation of energy and angular momentum via GWs is the process that ultimately drives the mergers we observe with LIGO and Virgo, this mechanism alone is not sufficient to explain the occurrence of coalescing compact binaries. Quasi-circular BH binaries of $\ssim 10 M_\odot$ at separations larger than $\ssim 10 R_\odot$ take more than a Hubble time to merge under GW radiation reaction alone. One thus needs some additional astrophysical mechanisms acting at larger separations. While it is well understood that individual BHs and NSs form from the collapse of massive stars,\cite{1986bhwd.book.....S} conceiving plausible astrophysical scenarios that can pair them in binaries while explaining the rates and properties of observed GW mergers still presents many open issues (see Refs.\cite{2018arXiv180605820M,2018arXiv180909130M} for recent reviews). Leading models of compact-binary formation include the isolated evolution of massive binary stars in galactic fields\cite{2014LRR....17....3P} via either common-envelope\cite{1976IAUS...73...75P}, stable mass transfer,\cite{2017MNRAS.471.4256V} or chemical mixing,\cite{2016MNRAS.458.2634M,2016A&A...588A..50M} and dynamical assembly aided by either a tertiary companion,\cite{2003ApJ...598..419W,2012ApJ...757...27A} multiple exchanges in dense clusters,\cite{2013LRR....16....4B} or gas-assisted migration.\cite{2017ApJ...835..165B}

Investigating the occurrence of hierarchical mergers constitutes an orthogonal, but at the same time complementary, direction to the usual ``field vs dynamics'' formation-channel debate.  The key question is the following: 
\begin{quote}
\it
What if, instead of being the direct product of stellar collapse, some of the BHs we observed with LIGO and Virgo are remnants of previous BH mergers?
\end{quote}
This line of research is often phrased as the reconstruction of the BH ``generation'', contrasting first-generation (1g) objects produced by stellar collapse to second- (2g) or higher- generation BHs involving remnants of previous mergers. In essence, a BH merger is a process that converts two parents into a single BH remnant and the emitted GWs.
Crucially, the properties of post-merger BHs (masses, spins, and proper velocities) are set by the relativistic dynamics of the merger process, not by their astrophysical environments. At the same time, however, assembling repeated mergers requires an astrophysical environment that efficiently retains post-merger remnants, such that they remain available to merge again.
  By relying largely on relativity, hierarchical mergers introduce a clean feature in the astrophysical modeling of GW events.

This reasoning opens the door to a number of intertwined questions. Does hierarchical assembly of BHs happen at all at the masses targeted by LIGO and Virgo? If yes, what is the fraction of hierarchically-formed events in the observable population of merging compact binaries? What does this constraint tell us about the branching ratios between the different BH binary formation channels?

This review is restricted to hierarchical mergers of stellar mass and their GW signatures. Among the LIGO/Virgo events, only a small fraction of systems are expected to be of hierarchical origin, with first-generation mergers comprising the bulk of the population. On the other hand, for supermassive BHs hosted at the center of galaxies, hierarchical assembly is believed to be the norm. The properties of BHs of $\ssim 10^{5}-10^{9} M_\odot$ are strongly correlated with those of their host environments\cite{2000ApJ...539L...9F} and are thus expected to grow through mergers as their host galaxies merge.\cite{1980Natur.287..307B} Supermassive-BH evolution has been comprehensively reviewed elsewhere.\cite{2010A&ARv..18..279V,2013ARA&A..51..511K,2014ARA&A..52..589H, 2017ogw..book...43C, 2020ARA&A..58...27I,2020arXiv201101994B} Hierarchical mergers of stellar-mass BHs, which are the focus of this review, have important synergies with the astrophysics of supermassive BHs, the former representing potential seeds for the latter.

This review is structured as follows. In Sec.~\ref{signatures}, we illustrate how hierarchical mergers provide characteristic GW signatures that are independent of the specific environment in which they form. In Sec.~\ref{formch}, we present the most promising astrophysical locations that can produce hierarchical mergers, which notably include star clusters and disks of active galactic nuclei (AGN). We stress that this paper is not designed to provide an exhaustive review of GW formation channels, but only to highlight their relevance to the hierarchical-merger problem. In Sec.~\ref{individual}, we review the interpretations of individual GW events as repeated mergers. In Sec.~\ref{pop}, we summarize ongoing efforts to statistically extract subpopluations of hierarchical mergers from the current catalog of LIGO/Virgo detections. Finally, we conclude and illustrate future prospects in Sec.~\ref{concl}.

\section{Gravitational-wave signatures}
\label{signatures}

Hierarchical mergers have unique GW signatures that make them distinguishable from BHs resulting from stellar collapse. Higher-generation BHs present, on average, \emph{both} larger masses and larger spins.

\subsection{Masses: populating the pair-instability mass gap}
\label{largemass}

Assembling multiple generations of BH mergers has the obvious effect of producing GW events with larger masses. The energy emitted in GW following a BH merger is $\lesssim 5\%$ of the total mass of the merging binary.\cite{2005PhRvL..95l1101P,2018PhRvD..98j4057O} BHs of higher generation inherit $\ssim 95\%$ the combined mass of their parents and are therefore heavier than both of them.

This point becomes particularly relevant in the context of the so-called ``pair-instability mass gap'' (sometimes also referred to as ``upper mass gap", in contrast with the putative ``lower mass gap'' between BHs and NSs). 
At the onset of stellar collapse, sufficiently large helium cores (in the mass range $\ssim30 - 130\,M_\odot$) reach central temperatures larger than $10^9 K$ at densities below $10^6 {\rm g/cm^3}$, triggering efficient electron-positron production. As radiation-pressure support drops, the adiabatic index in the core falls below $4/3$, causing a contraction of the core. This, in turn, ignites explosive burning of carbon and oxygen, producing an amount of energy that is comparable to the binding energy of the star. Depending on the mass of the collapsing core, the star can be either completely (``pair-instability supernova'')\cite{2003ApJ...591..288H} or partially (``pulsational pair-instability supernova'')\cite{2007Natur.450..390W} disrupted. 
This creates the lower edge of the pair-instability gap, an upper 
limit on the 
BH mass that can be produced following stellar collapse.\cite{2016ApJ...824L..10W,2016A&A...594A..97B,2019ApJ...882..121S,2020MNRAS.497.1043D} 
In stars with even more massive helium cores ($\gtrsim 130 M_\odot$), some of the energy from the core contraction goes into photodisentegration, preventing complete stellar disruption. If such massive stars exist, this would result in a population of BHs ``above the gap'' with masses $\gtrsim 120 M_\odot$.\cite{2019ApJ...883L..27M,2020arXiv200602211M }

Farmer \etal\cite{2019ApJ...887...53F,2020ApJ...902L..36F} found that pair instability creates a mass gap in the BH mass spectrum starting at $m\simeq 45 M_\odot$. This threshold was found to be a relatively solid prediction given the current understanding of stellar evolution. One of the largest uncertainties in the stellar evolution models is given by the unknown $^{12}{\rm C}(\alpha, \gamma)^{16}{\rm O}$ reaction rate, which may shift the pair-instability BH mass threshold between $\ssim40 M_\odot$ and $\ssim 55 M_\odot$ (up to $\ssim 90 M_\odot$ is one allows for variations at the 3-$\sigma$ level). Similarly, Renzo \etal\cite{2020MNRAS.493.4333R}  found a threshold of $48 M_\odot$, robust against the treatment of convection in the stellar evolutionary models.
Additional pieces of the puzzle include potential correlations of the location of the pair-instability gap with the BH spin\cite{2020A&A...640L..18M,2021arXiv210307933W} which might the impact edge of the gap by $\ssim 15\%$, uncertainties in current stellar-wind prescriptions,\cite{2020ApJ...890..113B, 2021MNRAS.504..146V} as well as dredge-up episode during the helium-burning phase\cite{2021MNRAS.501.4514C} which can also push the lower edge of the mass gap all the way to $\ssim 90 M_\odot$. 
GW observations from the first two observing runs of Advanced LIGO and Virgo provided observational evidence for a dearth of BHs heavier  than $\ssim 45 M_\odot$,\cite{2017ApJ...851L..25F,2019ApJ...882L..24A,2020arXiv201014533T,2020PhRvD.102l3022R} widely thought to be a consequence pair instabilities in supernovae. This observation offered the first empirical constraints on the pair-instability gap as well as a chance to test the underlying stellar and nuclear astrophysics.
 
The pair-instability  process immediately translates into a promising signature of hierarchical mergers in GW observations: if BHs with $50 M_\odot \lesssim m \lesssim 120 M_\odot$ cannot be produced by stars, they might well be the remnants of previous BH mergers. Possible caveats to this statement include envelope retention in low-metallicity population III stars,\cite{2021ApJ...910...30T,2021MNRAS.502L..40F, 2021MNRAS.501L..49K} stellar mergers prior to BH formation,\cite{2020ApJ...904L..13R,2020MNRAS.497.1043D,2020ApJ...903...45K,2021ApJ...908L..29G} evolution in detached binaries,\cite{2021arXiv210307933W} as well as accretion in either molecular clouds,\cite{2021ApJ...908...59R} minihalos,\cite{2020ApJ...903L..21S} dense clusters,\cite{2019A&A...632L...8R,2021MNRAS.501.1413N} or isolated binaries.\cite{2020ApJ...897..100V}

\subsection{Spins: a characteristic value}
\label{largespins}

Spins of hierarchically-formed BHs also present a distinctive feature. As two BHs merge, the sum of their component spins $\mathbf{S}_i$ and the orbital angular momentum at plunge $\mathbf{L}$  is converted into the spin of the final BH: $\mathbf{S}_{\rm fin}\simeq \mathbf{L} + \mathbf{S_1} + \mathbf{S_2}$. For the simpler case of an equal-mass, non-spinning merger, the remnant BH has a dimensionless Kerr parameter $S/ m^2 \equiv \chi \sim 0.69$.\cite{2005PhRvL..95l1101P,2009PhRvD..79b4003S} Somewhat surprisingly, this remains true in a statistical sense for generic BH mergers.\cite{2008ApJ...684..822B,2017PhRvD..95l4046G,2017ApJ...840L..24F,2021CQGra..38d5012G} This property of relativistic dynamics can be explained heuristically using the so-called ``orbital hang-up" effect.\cite{2006PhRvD..74d1501C} When BH spins are co-aligned with the orbital angular momentum, the binary inspirals for longer before merger. Their delayed plunge decreases the magnitude of $\mathbf{L}$, but spins and angular momentum add constructively into the spin of the remnant, $S_{\rm fin}\simeq L+ S_1+S_2$. On the other hand, binaries with anti-aligned spins plunge from afar: the orbital angular momentum at plunge is larger, but it is partially cancelled by the component spins,  $S_{\rm fin}\simeq L- S_1 - S_2$. In practice, these two effects counterbalance each other and produce a distribution of remnant spins that is strongly peaked at the characteristic value $\chi\simeq 0.7$.

Such BH spin values are thought to be  larger than those that can be produced by stellar collapse. Recent models of core-envelope interactions in massive stars point towards an efficient transfer of angular momentum out of the collapsing core,\cite{2019ApJ...881L...1F} resulting in BHs with $\chi\sim 10^{-2}$. This finding is partially supported by current GW data, which points to a sizable population of BHs with relatively small spins\cite{2019ApJ...882L..24A,2020arXiv201014533T,2020PhRvD.102l3022R,2020ApJ...895..128M} (but see also Refs.\cite{2017PhRvL.119y1103V,2020arXiv200709156B}). Tidal interactions can also play a relevant role: for binaries formed in isolation, tides have the net  effect of increasing the spin magnitudes and aligning the spin directions with the binary orbital angular momentum.\cite{1981A&A....99..126H,2018A&A...616A..28Q,2018PhRvD..98h4036G,2018MNRAS.473.4174Z,2020A&A...635A..97B,2021PhRvD.103f3032S}

If, overall, stars produce BHs which are  slowly rotating, the high-spin region of the parameter space around $\chi\simeq 0.7$ becomes exclusive to hierarchical mergers.\cite{2017PhRvD..95l4046G,2017ApJ...840L..24F,2020ApJ...894..129S} Baibhav \etal\cite{2020PhRvD.102d3002B} referred to this peculiarity of hierarchical mergers as the ``spin gap'', in analogy with the mass gap described above.

\begin{figure*}[t]
\centering 
\includegraphics[width=\textwidth]{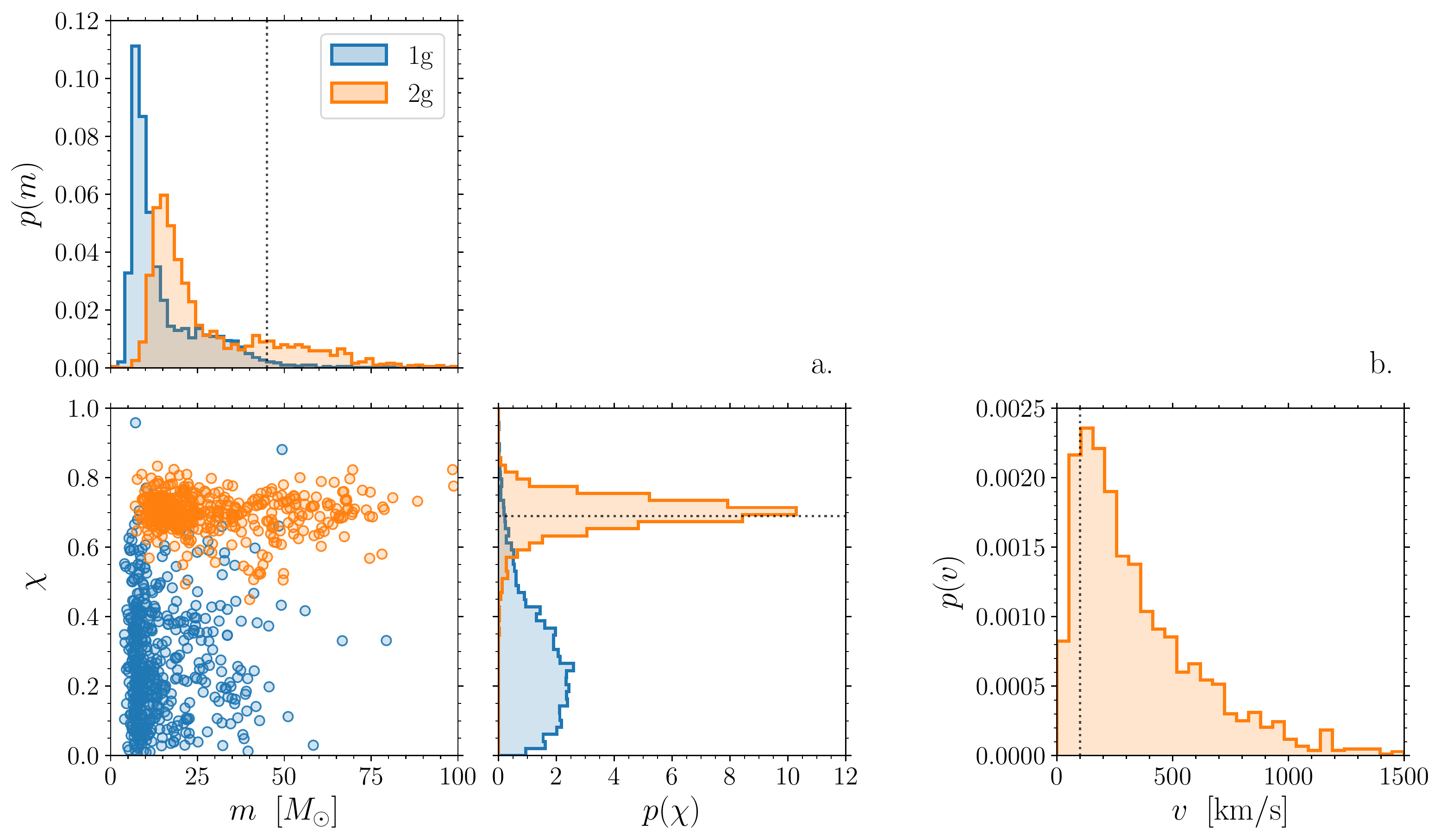}
\caption{\textbf{Masses, spins, and recoil velocities of first- and second-generation BHs}. The corner plot on the right (panel a.) shows BH masses $m$ and spins $\chi$. The histogram on the left (panel b.) shows the corresponding kick velocities $v$. Blue scatter points and histograms indicate a population of 1g BHs extracted from current LIGO/Virgo population fits.\cite{2020arXiv201014533T} Orange scatter points and histograms indicate the corresponding distribution of their merger remnants, which might form 2g GW events. 
Black dotted lines indicate typical values of astrophysical relevance: (i) the edge of the pair-instability mass gap\cite{2019ApJ...887...53F} $m=45 M_\odot$, (ii) the remnant spin of equal-mass, non-spinning BH mergers\cite{2009PhRvD..79b4003S}  $\chi = 0.69$, and (iii) an approximate upper limit to the escape speed of globular clusters\cite{2004ApJ...607L...9M,2002ApJ...568L..23G} $v=100$ km/s.%
}
\label{simplegens}
\end{figure*}

\subsection{A simple realization}
\label{simplerel}

The left panels of Fig.~\ref{simplegens} show a simple implementation of these ideas. 
We start from a population of 1g BHs modeled after current LIGO/Virgo detections\cite{2020arXiv201014533T} (thus implicitly assuming, for concreteness in this example, that all current events are of first generation). In particular, we consider their ``power-law + peak'' mass model and ``default'' spin model, which were found to return the highest Bayesian evidence among various phenomenological options. We extract binaries from the resulting posterior population distribution and estimate the properties of their merger remnants using fits to numerical-relativity simulations by Varma \etal\cite{2019PhRvR...1c3015V,2019PhRvL.122a1101V} The fits are evaluated at the reference GW frequency of $20$ Hz. Because the LIGO/Virgo population model only captures masses, spin magnitudes and polar spin angles, we distribute the azimuthal spin angles uniformly, which is equivalent to the prior used in the underlying single-event analyses.
About $0.1\%$ of the samples have mass ratios smaller than 1:6, which is outside the validity range of the remnant numerical-relativity fits.\cite{2019PhRvR...1c3015V} For those few cases, we implement expressions for final mass,\cite{2012ApJ...758...63B} spin,\cite{2016ApJ...825L..19H} and recoil\cite{2016PhRvD..93l4066G} which explicitly include the test-particle limit.  See Doctor \etal \cite{2021arXiv210304001D} for a study dedicated to the distribution of post-merger remnants extrapolated from the current LIGO/Virgo detections.

In this data-driven 1g population, only $\ssim 2\%$ of the BH masses (including primary and secondary components) are $>45 M_\odot$. This fraction goes up to $\ssim 20\%$ if one considers the 2g merger remnants,  indicating that the hierarchical mergers are indeed an efficient way to populate the mass gap. The 1g spin distribution extracted from current LIGO/Virgo data shows a broad preference for $\chi\lesssim 0.5$ (although some effective spin parameters are better measured).\cite{2020arXiv201014533T} The spin magnitudes of 2g BHs, however, are highly concentrated close to $\chi\ssim 0.7$.

 In general, repeated mergers are expected to present a strong mass-spin correlation and cluster in the high-$m$ high-$\chi$ region of the parameter space. This is a distinctive  population feature that might aid the identification of hierarchical mergers from LIGO/Virgo data (cf. Sec.~\ref{pop})

\subsection{Other observables}

The mass ratio of merging BHs might also provide precious insights, especially if different generations of mergers are combined. Very simply, one can expect that a mixed binary will present a mass ratio more unequal  compared to other events where both components come from the same generation. For instance, the mass ratio of a 1g+2g event might be somewhat close to $1/2$.

The redshift distributions of merging BH binaries could also contain information about their generation. Some delay will be present between the formation of first- and second-generation BHs which implies that the redshift of the hierarchical merger event should be lower than those of the previous generation.\cite{2017PhRvD..95l4046G} Specific astrophysical environments can, however, overpopulate the low-redshift region for 1g BHs. For instance, this might be the case for binaries that are ejected from clusters and take a long time to merge under gravitational radiation reaction, compared to binaries that merge relatively quickly inside the cluster. One also needs to consider that the higher masses involved in hierarchical mergers imply that larger redshifts are accessible to the detectors.

\begin{figure*}[t]
\centering 
\includegraphics[width=0.8\textwidth,trim=0.1cm 4.9cm 0.8cm 5.4cm,clip]{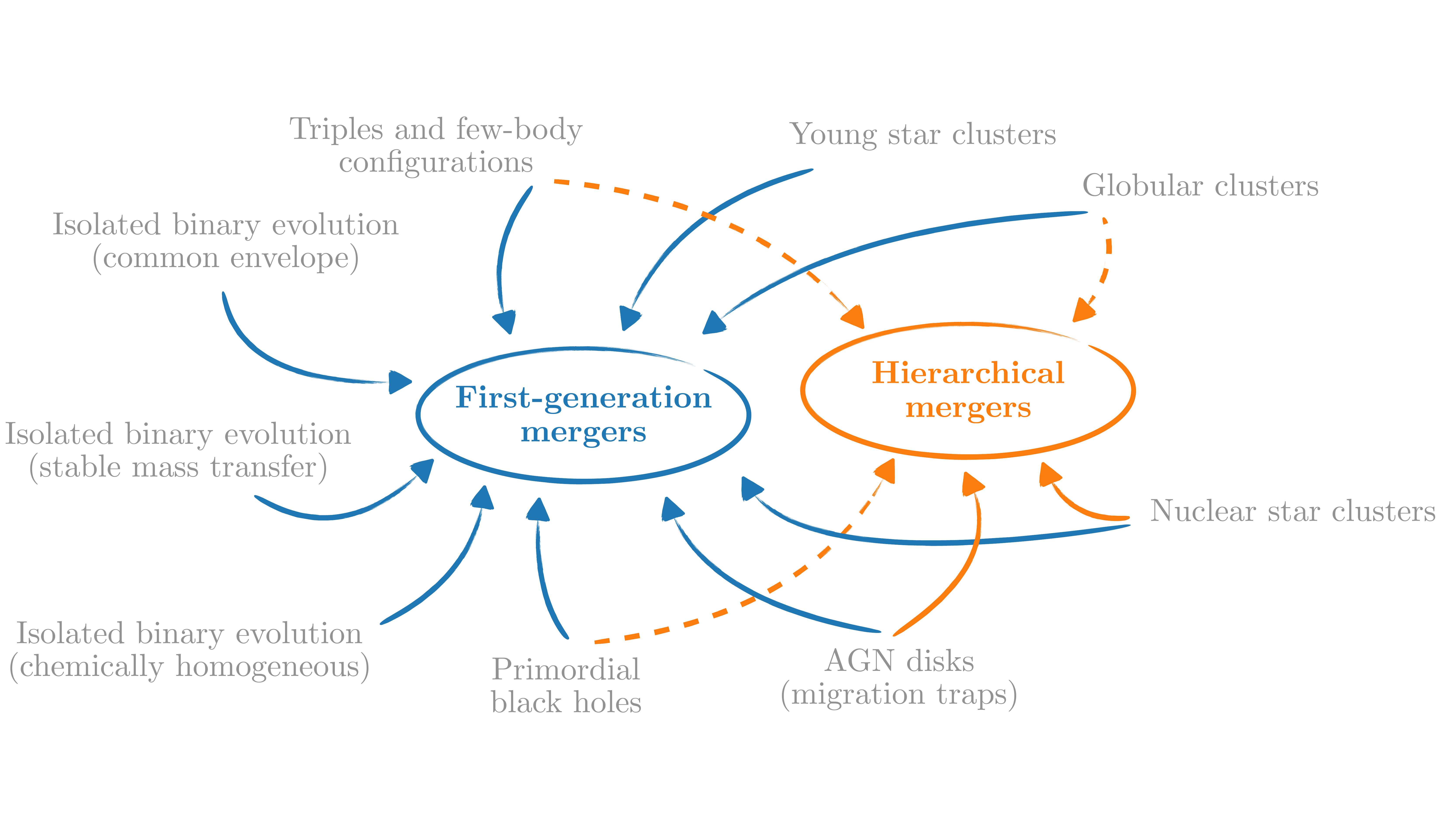}
\caption{\textbf{Interplay between the occurrence of hierarchical mergers and (some of) the proposed formation channels of merging compact binaries.} While all channels naturally produce 1g BHs (blue) either from stellar collapse or cosmological perturbations, only a subset of them can efficiently assemble hierarchical mergers (orange). %
Solid (dashed) lines indicate formation channels that are efficient (inefficient) at producing hierarchical mergers, though their overall contribution to the observed population will also depend on the relative merger rates from the different channels.}
\label{hmerdiagram}
\end{figure*}

\section{Astrophysical formation channels}
\label{formch}

Understanding the formation channels of merging BH binaries is now one of the most pressing quests in high-energy astrophysics. Among the many proposed scenarios, two main classes of models can be identified. In the first class, merging BHs originate from isolated binary stars. Alternatively, BH mergers can be aided by interactions with external bodies, such as other compact objects or large gaseous structures.   
As schematically summarized in Fig.~\ref{hmerdiagram}, hierarchical mergers are a prerogative of this second class of models.

\subsection{The role of the escape speed}

The escape speed of the environment has a crucial role in the assembly of hierarchical mergers.\cite{2019PhRvD.100d1301G} Second-generation BH binaries can be produced only if the remnants of previous mergers are efficiently retained within their astrophysical host.

As two BH merge, asymmetric dissipation of linear momentum via GWs imparts a recoil (or ``kick") to the post-merger remnant. These merger recoils range from 0 (for the case of highly symmetric configurations)\cite{2008PhRvL.100o1101B} to $\ssim 5000$ km/s (for the case of  equal-mass systems with large, misaligned spins).\cite{2007PhRvL..98w1102C,2007PhRvL..98w1101G,2011PhRvL.107w1102L}
For sufficiently broad populations,\cite{2007ApJ...662L..63S,2012PhRvD..85h4015L,2018PhRvD..97j4049G} BH merger kicks are of $O(100)$ km/s. For instance, using the 1g population extracted from  current LIGO/Virgo data as in Sec.~\ref{simplerel}, one finds that $\ssim 99\%$, $\ssim 85$\%, and $\ssim 3\%$ of the resulting 2g BHs receive a kick greater than $30$km/s, $100$ km/s, and $1000$ km/s, respectively.

In order to produce hierarchical mergers, the escape speed of the host needs to be larger than the typical kick imparted to BH remnants. 
For context, the escape speed of a typical globular cluster is $\ssim10-100$ km/s, that of the Milky Way is $\ssim 600$ km/s, while large elliptical galaxies can reach escape speeds of $\ssim 2000$ km/s.\cite{2002ApJ...568L..23G,2004ApJ...607L...9M}
If all dynamical formation channels were to produce BH mergers at approximately the same rate, it would be natural to expect that hierarchical mergers are more likely to originate from environments with larger escape speeds.
For the formation channels highlighted below, current BH-binary merger-rate predictions range from $4-60  \, {\rm Gpc}^{-3} {\rm yr}^{-1}$ (at $z\!=\!0$) for globular clusters,\cite{2018ApJ...866L...5R,2018PhRvL.121p1103F} $\ssim 1 \, {\rm Gpc}^{-3} {\rm yr}^{-1}$ for nuclear star clusters,\cite{2009ApJ...692..917M,2016ApJ...831..187A} and  $0.002-60 \,{\rm Gpc}^{-3} {\rm yr}^{-1}$ for AGN disks.\cite{ 2020A&A...638A.119G,2020ApJ...898...25T}

It is worth noting that orbital eccentricities also affect merger recoils. For eccentricities $\lesssim 0.3$,  BH kicks can be enhanced by up to $25\%$,\cite{2007ApJ...656L...9S,2020PhRvD.101b4044S,2021arXiv210111015R} making it more difficult to retain the remnants of mergers formed in dynamical environments, a fraction of which is predicted to be eccentric.\cite{2003ApJ...598..419W, 2012ApJ...757...27A,2018ApJ...860....5G,2018PhRvD..97j3014S,2019ApJ...871...91Z,2020arXiv201102507G} 
The consequences of eccentricity-enhanced kicks on the rate of hierarchical mergers is unexplored.

\subsection{Star clusters}

Repeated mergers in stellar cluster have long been investigated as potential formation pathways of intermediate-mass\cite{1989ApJ...343..725Q,1993ApJ...418..147L,1999A&A...348..117P,2001ApJ...562L..19E,2002ApJ...566L..17M,2002MNRAS.330..232C,2002ApJ...576..899P,2004ApJ...616..221G,2006ApJ...640..156G,2008ApJ...686..829H,2015MNRAS.454.3150G,2016MNRAS.459.3432M,2018ApJ...856...92F,2018PhRvD..97l3003K} and supermassive\cite{1987ApJ...321..199Q,2011ApJ...740L..42D,2001ApJ...558..535M,2003ApJ...593..661V,2014MNRAS.442.3616L} BHs.
In the LIGO context, an early report of potential hierarchical mergers was made by O'Leary \etal,\cite{2016ApJ...824L..12O} who claimed that such events could constitute up to $\ssim 10\%$ of the GW mergers from dense clusters (see e.g. their Fig.~2). Rodriguez \etal\cite{2018PhRvL.120o1101R,2019PhRvD.100d3027R} later presented a wide range of Monte Carlo integrations specifically targeting repeated mergers in globulars. They identified a very strong dependence on the birth spins of BHs: the contribution of hierarchical mergers to the BH merger rate drops from $\ssim 10\%$ if BHs are born non-spinning to $\lesssim 1\%$ for moderate spins $\chi\sim 0.5$. This is because higher-spinning BHs are subject to larger merger recoils and are thus more easily ejected. The simplified model by Gerosa and Berti\cite{2019PhRvD.100d1301G} also finds very small fractions of hierarchical mergers from globulars if spinning BHs are considered.

Semi-analytical treatments based on simulated stellar populations\cite{2020ApJ...894..133A,2020arXiv200715022M,2021arXiv210305016M} suggest that the fraction of repeated mergers in nuclear star cluster is $\ssim 1$ order of magnitude larger than that of globulars and $\ssim 3$ orders of magnitude larger than that of young star clusters. %
Similarly, energy arguments that relate the hardening rate of BH binaries to the global properties of the clusters\cite{2019MNRAS.486.5008A} indicate that the occurrence of repeated mergers presents a steep increase in systems with escape speeds $\gtrsim 300$ km/s and mass densities $\gtrsim 10^5 M_\odot/{\rm pc^3}$.
 Monte Carlo simulations\cite{2018MNRAS.481.2168M} and  further analytical modeling\cite{2020arXiv200609744S,2020PhRvD.102d3002B,2021MNRAS.502.2049L,2020MNRAS.498.4591F} produce qualitatively similar results: populating the upper mass gap via in-cluster GW mergers seems possible, but requires sufficiently massive environments.

Overall, these findings point towards galactic nuclei\cite{2009ApJ...692..917M,2009MNRAS.395.2127O, 2015MNRAS.448..754H,2016ApJ...831..187A} as the most likely cluster environments to host repeated mergers. The key binary formation mechanism is different for nuclear star cluster that do or do not host a central supermassive BH.\cite{2010IAUS..266...58B} In the former case, short relaxation time can result in the formation of a steep density cusp of stellar-mass BHs around the central supermassive BH, which facilitates mergers by GW captures.\cite{2009MNRAS.395.2127O} On the other hand, nuclear star clusters without a supermassive BH are akin to heavier globulars where hardening is driven by three-body encounters.\cite{2016ApJ...831..187A}

Nevertheless, it is important to point our that globular clusters could also host a sizable population of second-generation mergers if BH spins at birth turn out to be small, which is in line with some of the current predictions.\cite{2019ApJ...881L...1F} Furthermore, globular clusters were on average $\ssim 5$ times more massive at birth compared to the present time,\cite{2015MNRAS.453.3278W} which increases their escape speeds by a factor of $\ssim \sqrt{5}> 2$. These are crucial details because globulars are thought to be extremely efficient factories of GW events.\cite{2010MNRAS.402..371B, 2013MNRAS.435.1358T,2015PhRvL.115e1101R, 2016PhRvD..93h4029R,2016ApJ...824L...8R,2017MNRAS.464L..36A,2017MNRAS.469.4665P,2020PhRvD.102l3016A}

The cluster metallicity might play an important role in the formation of hierarchical GW events, with a preference for metal-poor environments\cite{2020arXiv200715022M} (but see Ref.\cite{2020ApJ...902L..26F} for opposite claims). Additionally, a notable boost in the rate of hierarchical stellar-mass BH mergers in clusters could be provided by Kozai-Lidov oscillation induced by a massive perturber.\cite{2021MNRAS.502.2049L} We also note that hierarchical mergers involving NSs have also been explored as a potential formation channel of GW events with one of more components in the lower mass gap ($3M_\odot \lesssim M \lesssim 5M_\odot$), both in clusters\cite{2020PhRvD.101j3036G}
 and few-body configurations in the field.\cite{2020ApJ...888L...3S,2021MNRAS.500.1817L,2021ApJ...907L..19V,2021arXiv210303782H}

\subsection{AGN disks}

Gaseous AGN disks are also promising environments for the production of BH binaries merging in the LIGO/Virgo band.\cite{2017ApJ...835..165B,2017MNRAS.464..946S,2018MNRAS.474.5672L, 2018ApJ...866...66M,2019ApJ...878...85S,2020A&A...638A.119G,2020ApJ...898...25T} In this scenario, stellar-mass BHs are embedded in accretion disks surrounding supermassive BHs, and their evolution is driven by angular-momentum transfer  via viscous interactions ---a process that is analogous to that of planetary migration in protoplanetary disks.\cite{2013apf..book.....A} Bellovary \etal\cite{2016ApJ...819L..17B} pointed out that semi-realistic disk models\cite{2003MNRAS.341..501S,2005ApJ...630..167T} predict the existence of ``migration traps.'' These are specific locations in the disk where the viscous torque changes signs: disk perturbers at radii larger than the trap migrate inwards, while those at smaller radii migrate outwards. Trapped migration is an ideal mechanism to assemble multiple generation of BH mergers: stellar-mass BHs either formed by disk fragmentation or captured by the gravitational pull of the AGN tend to migrate toward the very same locations and thus meet each other. Orbital velocities at the migration traps are of $\mathcal{O}(10^4)$ km/s,\cite{2016ApJ...819L..17B} which makes these systems largely insensitive to BH merger kicks of $\ssim 100$ km/s.

Hierarchical mergers in AGN disks have been explored, with focus on both intermediate mass\cite{2012MNRAS.425..460M,2014MNRAS.441..900M} and stellar-mass events detectable by LIGO/Virgo.\cite{2018ApJ...866...66M,2020MNRAS.494.1203M,2019PhRvL.123r1101Y,2020ApJ...898...25T} Assuming that (i) the number of BHs  dragged into a migration trap within the AGN lifetime follows a Poisson distribution and (ii) once a new BH reaches the trap it mergers immediately with the BH that is already there, Yang \etal\cite{2019PhRvL.123r1101Y} found that the fraction of higher-generation mergers from AGN disks is $\gtrsim 50\%$. Combining N-body simulations and analytical arguments, Tagawa \etal\cite{2020ApJ...898...25T} reports fractions between $\ssim 20\%$ and $\ssim 45\%$, although in their models migration traps are less relevant to the overall merger rate. For BHs formed in AGN disks, repeated mergers are a very likely outcome, perhaps even the most likely.

Unique signatures of repeated mergers formed in AGN disks include not only large masses (Sec.~\ref{largemass}) and large spin magnitudes (Sec.~\ref{largespins}), but also preferential spin alignment. This process is commonly refereed to as the Bardeen-Petterson effect\cite{1975ApJ...195L..65B} and has long been explored for the case of supermassive BH binaries.\cite{2007ApJ...661L.147B,2013ApJ...774...43M,2016MNRAS.460.1243R,2020MNRAS.496.3060G}  Some gas from the AGN disc will form a smaller  circumbinary disk surrounding the two stellar-mass BHs which, in turn, feeds individual disks around each of the two components. General-relativistic frame dragging excites warps in these ``minidisks''\cite{2017ApJ...835..199R} which tend to co/counter-align the BH spins with the disk's angular momentum. Assuming that the binary's orbital plane and the AGN disk are coplanar, the occurrence of the Bardeen-Petterson effect implies some degree of preferential alignment between the spins and the binary's angular momentum, with potential GW signatures.  
 
This argument has recently been made explicit 
by McKernan \etal\cite{2020MNRAS.494.1203M} Crucially, the Bardeen-Petterson effect allows for both co- and counter-alignment.\cite{2005MNRAS.363...49K} This is because the disk constitute an essentially axisymmetric environment. Each of the two spins of a BH binary can thus be either co- or counter-aligned with the orbital angular momentum which implies the presence of four, roughly equally populated subpopulations (``up-up'', ``up-down'', ``down-up'', and ``down-down''). Notably, three of these will still be close to their aligned configuration as they enter the LIGO band, while ``up-down'' binaries, in which the more (less) massive BH is co- (counter-) aligned, are expected to precess away because of a dynamical instability.\cite{2015PhRvL.115n1102G,2016PhRvD..93l4074L,2020PhRvD.101l4037M,2021PhRvD.103f4003V} Counter-alignment thus distinguishes the AGN scenario from both isolated binaries (where only co-alignment is predicted) and dynamical assembly in clusters (where spin orientations are randomized by frequent encounters), constituting a potential smoking-gun signature of disk-driven mergers.

Because of their gaseous environment, GW events from BHs in AGN disks could present electromagnetic counterparts. Super-Eddington accretion,\cite{2017ApJ...835..165B} as well as re-adjustment of the Hill sphere  due to GW mass loss and recoil\cite{2019ApJ...884L..50M,2020PhRvL.124y1102G} are expected to produce detectable flares. Suggestive associations with fast radio bursts (FRBs) have also been proposed.\cite{2019ApJ...884L..12Y}

\subsection{Exotica}

An entertaining possibility is that  BH mergers detectable by LIGO/Virgo have a cosmological origin and constitute (a fraction of)  the Universe's dark matter\cite{2016PhRvL.116t1301B} see e.g.~Ref.\cite{2020MNRAS.496..994K} for detection-rate estimates and Refs.\cite{2020PhRvD.102l3524H,2021arXiv210203809D} for model selection).

The impact of second-generation mergers on the predicted population of such primordial BHs was explored in Refs.\cite{2019EPJC...79..717L,2020PhRvD.101h3008W,2020JCAP...04..052D}. De Luca \etal\cite{2020JCAP...04..052D} found that the overall contribution of repeated mergers to the primordial BH population is $\lesssim 0.5\% \times (f_{\rm PBH}/10^{-3})^{16/37}$, where $f_{\rm PBH}$ is the fraction of primordial BH in dark matter. This estimate depends very mildly on the assumed mass spectrum: it thus seems unlikely that the merger history of primordial BHs could play a relevant role in the LIGO/Virgo context (but see e.g. Ref.\cite{2015PhRvD..92b3524C} for a different result).

Bianchi \etal\cite{2018arXiv181205127B} considered an interpretation of GW observations in quantum gravity, arguing that BHs are in micro-canonical equilibrium (i.e. at fixed energy and particle number) and might undergo repeated mergers. In this scenario, the Bekenstein-Hawking entropy formula predicts minuscule spin magnitudes ($\chi\ssim 10^{-38}$ for BHs heavier than a Planck mass), while the high-spin region of the parameter space is populated exclusively  by merger remnants.

\section{Individual-event constraints}
\label{individual}

The first three observing runs of LIGO/Virgo saw a number of exceptional systems, including very massive BHs or significantly asymmetric binaries.
Several authors have speculated whether some of these systems may be products of hierarchical assembly. 

\subsection{GW170729}

With a masses of $\ssim50M_\odot+35 M_\odot$ and mild evidence for spin precession, GW170729 is
the heaviest system reported in the GWTC-1 catalog from the first and second  LIGO/Virgo observing runs.\cite{2019PhRvX...9c1040A} The possibility that GW170729 contains a 2g BH was approached by tuning the Bayesian prior,\cite{2019PhRvD.100j4015C} statistical comparisons with the other GWTC-1 event,\cite{2020ApJ...893...35D,2020RNAAS...4....2K,2020ApJ...900..177K} and more specifically in the context of BH formed in clusters\cite{2020ApJ...894..133A} and AGN disks.\cite{2019PhRvL.123r1101Y} These studies found only weak preference in favor of a hierarchical interpretation indicating that, overall, data are inconclusive. Refs.\cite{2019ApJ...882L..24A,2020ApJ...891L..31F} both argue that masses and spins of GW170729 are consistent with the other GWTC-1 events, suggesting that it is not  necessary to invoke a hierarchical origin to explain this system.

\subsection{GW170817A} 
The independent re-analysis of data from first two LIGO/Virgo observing runs by Zackay \etal\cite{2019arXiv191009528Z} reported the additional trigger GW170817A, which is not present in GWTC-1 (not to be confused with the more famous NS merger GW170817\cite{2017ApJ...848L..12A} from the same calendar day). If this event is of astrophysical origin (which is unclear\cite{2019arXiv191009528Z, 2020arXiv200800509P}) %
its total mass of $\ssim 100 M_\odot$ and effective aligned spin parameter of $\chi_\mathrm{eff} \sim 0.5$ makes it a promising candidate for a 2g merger. According to the models of  Gayathri \etal,\cite{2020ApJ...890L..20G} a repeated merger in an AGN disk should be preferred compared to an isolated origin. The event remains consistent with component masses below $50\,M_\odot$.

\subsection{GW190412} 
\label{massratioguy}
The first reported  system with confidently unequal masses is GW190412, with a mass ratio of $q \sim 0.3$.\cite{2020arXiv200408342T} This is in contrast to the 10 BH events from GWTC-1, which were all consistent with $q = 1$ and suggested a strong preference for equal mass mergers and a median mass ratio of $\ssim 0.9$.\cite{2020ApJ...891L..27F} Most of the GWTC-2 events\cite{2020arXiv201014527A} are also consistent with $q=1$.  GW190412 exhibits a weak preference for spin precession caused by component BH spins that are misaligned with the orbital angular momentum. %
 While measurements of some effective spin quantities\cite{2001PhRvD..64l4013D,2008PhRvD..78d4021R,2015PhRvD..91b4043S,2021PhRvD.103f4067G} are solid, identifying the relative contribution of the two component spins is more challenging. The LIGO/Virgo analysis\cite{2020arXiv200408342T} found that the primary BH is mostly responsible for the spin signature in the data. Mandel and Fragos\cite{2020ApJ...895L..28M} argued that, because of tidal spin-up in isolated binaries, one should rather prefer a spinning secondary. Follow-up parameter-estimation studies\cite{2020ApJ...899L..17Z} indicate that the former interpretation should be preferred.
 
Hierarchical mergers provide a viable scenario to explain  both the asymmetric masses and the spin properties. Gerosa \etal\cite{2020PhRvL.125j1103G} showed that a 1g+2g interpretation of GW190412 returns similar Bayesian odds compared to a standard 1g+1g analysis but favors an environment with escape speed $v_{\rm esc}\gtrsim 150$ km/s, thus excluding globular clusters. In those models, a large escape speed is required to accommodate the measured values of the mass ratio and the effective spins.
Rodriguez \etal\cite{2020ApJ...896L..10R} interpreted the event as a 1g+3g merger in a super star cluster, which naturally explains the $\ssim 1/3$ mass ratio. GW190412 was also interpreted in the context of disk-driven formation\cite{2020ApJ...899...26T,2021ApJ...908..194T} and as the end product of a quadrupole star system.\cite{2020ApJ...898...99H}
However, one cannot exclude that this system belongs to the tail of the mass-ratio distribution, rather than a different formation channel\cite{2020arXiv200408342T} and, indeed, the more recent analyses of Refs.\cite{2020arXiv201014533T,2020arXiv201105332K} did not single GW190412 out as an outlier. Olejak \etal\cite{2020ApJ...901L..39O} argues that the mass ratio of GW190412 can be predicted by isolated common-envelope evolution, where that $\ssim 10\%$ of isolated, 1g binary BH systems might have $q<0.4$. The isolated channel has a more difficult time producing systems with significant spin precession (but see Refs.\cite{2000ApJ...541..319K,2018PhRvD..98h4036G,2021PhRvD.103f3032S}).

\subsection{GW190521}
\label{bigguy}
  The heaviest binary BH system observed thus far is GW190521, with a total mass of $\ssim 150\,M_\odot$.\cite{2020PhRvL.125j1102A} The primary mass of GW190521 is confidently above $65\,M_\odot$, placing it  inside the pair-instability mass gap and making GW190521 the most promising hierarchical merger candidate to date. A repeated merger origin is arguably considered to be the leading explanation of this event\cite{2020ApJ...900L..13A,2020arXiv201105332K}--- a conclusion that is further strengthened by evidence of spin precession\cite{2020PhRvL.125j1102A} and/or residual eccentricity.\cite{2020ApJ...903L...5R,2020arXiv200905461G,2020arXiv200901066C} These features support a dynamical origin for GW190521, which is a prerequisite for hierarchical formation (see also Refs.\cite{2020ApJ...902L..26F,2020arXiv201006161A} for further modeling in this direction).   The statistical mixture-model analyses of Abbott \etal\cite{2020ApJ...900L..13A} ended up favoring a 1g+1g model for GW190521, but this is because the underlying distributions from Kimball \etal\cite{2020ApJ...900..177K} are specifically tuned to globular clusters, which present escape speed lower than the typical merger kicks (cf. Sec.~\ref{formch}).  Subsequent work from the same group\cite{2020arXiv201105332K} classified GW190521 as a 2g merger but boosted the cluster escape speed to $\gtrsim 100$ km/s.
Specifically in the context of AGN disks, Tagawa \etal\cite{2021ApJ...908..194T} finds that GW190521 is compatible with a 2g+2g (higher-g) BH merger in metal-poor (metal-rich) environments.

 In addition to the GW constraints, the electromagnetic flare ZTF19abanrhr from AGN J124942.3+344929 detected\cite{2020PhRvL.124y1102G} by the Zwicky Transient Facility\cite{2019PASP..131a8002B} presents suggestive,\cite{2020arXiv200914057C} but far from conclusive,\cite{2020arXiv200912346A,2021arXiv210316069P} association with GW190521. The interpretation put forward by Graham \etal\cite{2020PhRvL.124y1102G} relies on shocks generated by a recoiling BH in the disk surrounding the AGN. A post-merger remnant of mass $m$ with kick velocity $v$ will influence a sphere of gas with radius $r\lesssim Gm/v^2$. The BH leaves bound gas behind after a time $t\sim r/v$ ($\ssim 30$ days for ZTF19abanrhr) and enters an unperturbed region of the disk. During this phase, the accretion rate becomes super-Eddington and the Bondi-Hoyle-Lyttleton luminosity reaches $\mathcal{O}(10^{45})$ ergs/s, providing the necessary energy to power the observed transient.\cite{2019ApJ...884L..50M,2020PhRvL.124y1102G} Interestingly, they predict that a similar flare should be detected after $\ssim 1.6$ yr as the kicked BH re-enters the disk. While it is not possible to rule out a chance coincidence for an AGN flare in the $\sim 4$ Gpc$^3$ 90\% localization volume of GW190521, which likely contains tens of thousands of AGNs, it remains an exciting possibility. The presence or absence of future associations between GW events and AGN flares will reveal whether or not there is a physical connection.\cite{2021arXiv210316069P} If confirmed, this interpretation would indicate that the host of GW190521 is an environment that can efficiently retain kicked remnants and thus host hierarchical mergers.

Several alternative explanations for the occurrence of GW190521 have also been put forward, ranging from critically assessing uncertainties in the lower edge of the mass gap,\cite{2020ApJ...905L..15B,2021MNRAS.501.4514C,2021MNRAS.502L..40F,2020ApJ...905L..21U} Population III stars at very low-metallicity,\cite{2021MNRAS.502L..40F,2021MNRAS.501L..49K,2020ApJ...903L..40L} accretion onto either stellar-origin\cite{2021ApJ...908...59R,2020ApJ...903L..21S,2021ApJ...908..194T} or primordial\cite{2021PhRvL.126e1101D, 2021arXiv210101705C} BHs, and stellar mergers.\cite{2020ApJ...904L..13R,2020MNRAS.497.1043D,2020ApJ...903...45K} Additional speculations include invoking beyond-standard-model physics\cite{2020arXiv200707889C,2020PhRvL.125z1105S} exotic compact objects,\cite{2020arXiv201005354A,2021PhRvL.126h1101B} and dark-matter annihilation.\cite{2020arXiv201000254Z} From a data-analysis perspective alone, there is also the concrete possibility that the primary and secondary components of GW190521 are actually above and below the pair instability gap respectively, and no object is inside it.\cite{2020ApJ...904L..26F,2021ApJ...907L...9N} %

\subsection{GW190814}
 While GW190521 contains a BH that appears to sit in the pair-instability mass gap, the event GW190814\cite{2020ApJ...896L..44A} is notable because one of its components lies in the purported \emph{lower} mass gap, between NSs and BHs.\cite{1998ApJ...499..367B,2010ApJ...725.1918O} The secondary mass of GW190814 at $\ssim2.6\,M_\odot$ is probably too large to be a NS,\cite{2020ApJ...896L..44A,2021ApJ...908L...1T} but rotation might be an key player to discriminate between the two outcomes.\cite{2021PhRvC.103b5808D,2020ApJ...905...48T,  2020MNRAS.499L..82M,2020arXiv201002090B} %
 It has been proposed that the empirical low mass gap, which was first observed by studying the masses of X-ray binaries, is caused by the supernova explosion mechanism,\cite{2012ApJ...749...91F,2012ApJ...757...91B} although selection biases might not be fully under control.\cite{2011ApJ...741..103F,2012ApJ...757...36K} If a different origin is really behind the existence of the low mass gap, the outlier mass of GW190814's secondary BH may indicate formation through a different pathway compared to the other events. At roughly the same mass as the merger product of GW170817,\cite{2017ApJ...848L..12A} the secondary GW190814 is suggestive of a hierarchical merger origin between NSs.\cite{2021MNRAS.500.1817L,2020PhRvD.101j3036G} However, NS mergers are predicted to be extremely rare in dynamical environments,\cite{2020ApJ...888L..10Y} which challenges this interpretation. In AGN disks, the interplay between repeated mergers and accretion has been suggested as a viable mechanism to produce GW190814-like events.\cite{2020ApJ...901L..34Y, 2021ApJ...908..194T}

\section{Model selection for populations of events}
\label{pop}

While individual events may hint at a hierarchical origin, robust constraints on the presence of hierarchical mergers in a GW catalog requires a population analysis. We refer the reader to Refs.\cite{2019PASA...36...10T,2020arXiv200705579V} for pedagogical introductions and Refs.\cite{2019ApJ...882L..24A,2020arXiv201014533T,2020PhRvD.102l3022R} for state-of-the-art applications. 

\subsection{Outliers and subpopulations}

The goal of population studies is to infer the collective properties ---masses, spins, redshifts, eccentricities---  from the detected set of binary BHs. If the observed population contains a subpopulation of systems formed through hierarchical mergers, jointly analyzing all of the events in the GW catalog can reveal the presence of this distinct set of events. Population analyses can provide powerful constraints on the astrophysical rate of hierarchical mergers and the natal masses and spins of BHs, even in the case where one cannot confidently identify which specific events in the catalog are of hierarchical origin.

The first signs of a new subpopulation may appear in the form of outliers in the data. As summarized in Sec.~\ref{individual}, a hierarchical merger origin was explored for some individual events because of their unusual properties. However, it is generally difficult to disentangle a genuine population outlier (for example, a binary BH system with a different origin) from a statistical fluctuation in the tails of the population, especially in the presence of large measurement uncertainties.\cite{2020ApJ...891L..31F} Furthermore, inference based on a single event can be highly dependent on the choice of parameter-estimation prior, at least at the present signal-to-noise ratios.\cite{2017PhRvL.119y1103V,2020ApJ...904L..26F,2020ApJ...899L..17Z} By considering a single outlier, the measurement uncertainties on its parameters, together with theoretical uncertainties on the potential subpopulation (from which there are $\leq 1$ observations), mean that one can rarely discern its origin with high confidence. On the other hand, combining information across multiple events might reveal whether there is indeed a subpopulation of BHs with the properties characteristic of hierarchical assembly.

As discussed in Sec.~\ref{signatures}, hierarchical mergers leave a distinct imprint on the spins of BHs. For individual events, BH spins are poorly measured, but, with $\mathcal{O}({100})$ events, a potential subpopulation with   $\chi \sim 0.7$  can be confidently identified in the data.\cite{2017ApJ...840L..24F,2017PhRvD..95l4046G} This method relies on simultaneously inferring, or knowing a priori, the spin distribution of first-generation BH binaries. If natal BH spins are very small,\cite{2019ApJ...881L...1F} it will take fewer observations to reveal the imprint of hierarchical mergers on the spin distribution. On the opposite extreme, if BHs are born with spins tightly clustered around $\chi \sim 0.7$, it will be difficult to infer the presence of hierarchical mergers from the spin distribution alone.\cite{2020PhRvD.102d3002B} 

In addition to imprinting spins, repeated mergers increase the masses of the merging BHs. Therefore, identifying a positive correlation between spins and  masses may point to the role of hierarchical mergers.\cite{2020ApJ...894..129S} Interestingly, some formation models\cite{2020A&A...636A.104B,2018PhRvD..98h4036G} predict a negative mass-spin correlation for 1g BHs which, if confirmed, might enhance the distinguishability of the 2g subpopulation. 
Using 47 events from GWTC-1 and GWTC-2, Abbott \etal\cite{2020arXiv201014533T} searched for evidence that the most massive BH mergers in the catalog follow a spin distribution that is distinct from the spin distribution of the lower-mass systems. This analysis modeled the binary BH primary mass distribution as a mixture of a power-law component with a high-mass Gaussian component, and allowed the distribution of spin magnitudes and spin tilts to vary between both components. While there are hints that the spin distribution within the high-mass Gaussian component favors larger spin magnitudes and larger spin tilts (approaching an isotropic tilt distribution), large uncertainties remain and there is no conclusive evidence that the spin distribution varies with mass. %
Additional phenomenological trends that can reveal the presence of hierarchical  mergers include a series of peaks\cite{2020arXiv201104502T} or a ``smoothed staircase'' structure\cite{2020ApJ...893...35D}  in the mass distribution, a forbidden region in the mass-spin plane,\cite{2021arXiv210411247G,2020A&A...640L..20B} and the flattening of the spin distribution at high masses.\cite{2021arXiv210409510T}
Measuring the evolution of the mass and/or spin distribution with redshift could also reveal subpopulations of higher generation mergers.\cite{2021arXiv210107699F}

Beside phenomenological models that search for trends in the BH component mass, mass ratio and spin distribution, data analysis strategies based on physical coagulation models have also been successfully employed.\cite{2018ApJ...858L...8C,2020ApJ...893...35D,2020arXiv200810389F,2020ApJ...900..177K} These methods typically parameterize the population of first-generation BHs with a phenomenological model and introduce additional parameters calibrated on cluster simulations to model the coagulation process and production of higher generation mergers. The resulting fit simultaneously measures the properties of the first-generation BH subpopulation and the higher-generation subpopulations. 
Both Doctor \etal\cite{2020ApJ...893...35D} and Kimball \etal\cite{2020ApJ...900..177K} find that the GWTC-1 catalog does not present evidence for hierarchical mergers. On the other hand, subsequent work by Kimball \etal\cite{2020arXiv201105332K} reports that the updated sample of GWTC-2 is likely to contain a subpopulation made of hierarchical merger products provided that the cluster escape speed is $\gtrsim 100$ km/s. Among the events, that which is more likely to be a repeated mergers is, perhaps unsurprisingly, GW190521 (c.f. Sec.~\ref{bigguy}). This finding is at odds with an earlier analysis by Abbott \etal\cite{2020ApJ...900L..13A} which relies on previous models\cite{2020ApJ...900..177K} and includes only the GWTC-1 events and GW190521. In the analysis of Kimball \etal,\cite{2020arXiv201105332K} the model with highest Bayes factor has $v_{\rm esc}\sim 300$ km/s, which is relatively far from the globular-cluster regime their models are tuned to. For this parameter choice, the events which might be of hierarchical nature beside GW190521 are GW190517, GW190519, GW190602, GW190620, and GW190706 ---a list that notably excludes GW190412 despite its mass ratio (cf. Sec.~\ref{massratioguy}).
Baxter \etal\cite{2021arXiv210402685B} explicitly introduced a 2g component in their population fit and found that it plays a marginal role.
 Veske \etal\cite{2021ApJ...907L..48V} attempted the more ambitious identification of pairs of events which may be directly related to each other. The search did not yield a statistically significant association between event pairs, the most likely relationship being GW190514 as the predecessor of GW190521.

Finally, another analysis strategy is to compare the data directly to a set of simulations, inferring unknown physical parameters in the simulation rather than introducing any phenomenological parameters.\cite{2018PhRvD..98h3017T,2021PhRvD.103h3021W,2021ApJ...910..152Z} This is currently challenging due to the large number of theoretical uncertainties, including but not limited to, the distribution of birth spins of BHs,\cite{2018MNRAS.473.4174Z,2019ApJ...881L...1F,2020A&A...636A.104B,2020A&A...635A..97B} which plays a key role in determining the retention fraction of BH merger products and therefore the expected rate of 2g mergers.\cite{2019PhRvD.100d1301G,2019PhRvD.100d3027R} Additionally, the large number of proposed formation channels makes it challenging to choose a particular set of simulations to compare against the data. Some work in this direction includes the development of importance-sampling algorithms\cite{2019MNRAS.490.5228B} and  machine-learning emulators\cite{2018PhRvD..98h3017T,2020PhRvD.101l3005W} to speed up the evaluation of the population likelihood.  A direct application of these ideas to the hierarchical-merger problem has, to be best of our knowledge, not yet been presented.

\subsection{Measuring the mixing fraction} 

The confident identification of one or more hierarchical mergers in the GW catalog will be an important step towards inferring the contributions of different BH-binary formation channels to the total merger rate. Although this measurements is a natural by-product of a complete population analysis which include multiple channels, it is worth sketching out the key principle here.    
  
Let us suppose for the sake of this argument that there are two channels at play, $A$ and $B$, such that each of them is responsible for a given fraction of GW events in the observed catalog: $f_{A}+f_{B}= 1$. This can be  be further separated into first-generation and hierarchical mergers, i.e. $f_{A, {\rm 1g}} + f_{A, {\rm 2g}}+ f_{B, {\rm 1g}} + f_{B, {\rm 2g}} = 1$. If a given theoretical model predicts that hierarchical mergers are possible with efficiencies $\lambda_i = f_{i, {\rm 2g}} / f_i$  (with $i=A,B$) and the fraction of hierarchical mergers $f_{\rm 2g} =  f_{A, {\rm 2g}} + f_{B, {\rm 2g}}$ can be identified experimentally from the data, one immediately obtains  the branching ratios  $f_{
A} = (\lambda_A - f_{2g}) / (\lambda_A -\lambda_B)$ and $f_{
B} = (\lambda_B - f_{\rm 2g}) / (\lambda_B -\lambda_A)$. The limit $\lambda_A\ll \lambda_B$ corresponds to a scenario where one of the two pathways is much more likely to produce repeated mergers, as in the case of isolated binary formation (A) and dynamical assembly (B). In this case, one immediately finds $f_B = 1-f_A \simeq f_{\rm 2g}/\lambda _B$. For instance, if GW190521 were the only 2g merger in GWTC-1 and GWTC-2 ($f_{\rm 2g}\sim 1/50$), a single dynamical channel that predicts $\lambda_B =10\%$ of repeated  mergers would indicate that $\ssim 10$ events should be dynamically assembled while the remaining $\ssim 40$ evolved in isolation.

\section{Conclusions}
\label{concl}

In the context of LIGO/Virgo observations, the hierarchical BH merger scenario attracted a steep increase of interest in the last few years. We have attempted a concise review of both observational constraints and modeling advances that underpin this progress.

LIGO and Virgo are sensitive to BHs with masses of $\ssim 10-100 M_\odot$, similar to those of stars. Stars have been long predicted and observed to leave behind compact objects at the end of their lives: the working assumption when interpreting GW data, therefore, is that the observed BHs are produced by stellar collapse. However,  both individual GW observations as well as the statistical properties of the entire detected population seem to indicate that stars might not be the progenitors of all LIGO events. Other BHs might have a role to play: some of the events LIGO/Virgo observes might not be the direct end-product of massive stars, but rather originate from previous BH mergers.  
Such repeated assembly is very reminiscent of supermassive BHs formation pathways, as those objects have long been predicted to grow hierarchically, tracking the formation of structure in our Universe. Current observations of stellar-mass compact objects by LIGO and Virgo might be giving us a glimpse of how the very first  few generations of hierarchical BHs are assembled. %

Notably, hierarchical BH mergers present key features that are set by the very fact that they are hierarchical, irrespectively of their precise astrophysical origin. First,
hierarchical BHs can evade the mass cutoff imposed by the pair-instability supernova process and populate the predicted upper mass gap. Second, the spin of hierarchical BHs is almost entirely determined by the relativistic emission of angular momentum at merger, a process which erases information of the spins inherited from the stellar progenitors. At the same time, BH remnants are imparted strong recoils at merger, which might eject them from their astrophysical host and prevent them from merging again. The key point is that all of these features (mass build up, remnant spin, and merger recoil) are set by the relativistic dynamics of the merger process. There is, therefore, the exciting possibility of identifying some of the LIGO/Virgo events as hierarchical, independently of the details of their specific formation pathway. That said, such identifications then need to be explained within astrophysical models able to accommodate repeated mergers in a consistent fashion. In particular, dynamical formation in nuclear star clusters and gas-assisted migration in AGN disks appear to be the most promising candidates. For these reasons, one should see the hierarchical assembly of BHs as an orthogonal but complementary strategy to constraining the formation channel(s) of merging compact binaries.

The LIGO, Virgo, and soon KAGRA interferometers are now on a solid path to deliver thousands of BH binary observations within a few years.\cite{2020LRR....23....3A} A population of stellar-mass hierarchical BH mergers, if it is truly there in the Universe as tentatively suggested by current data, is bound to emerge with increasing clarity. In the coming years, the study of hierarchical BH mergers of stellar origin and their GW signatures will be a continuous dialogue between theoretical modeling and observational findings, each field repeatedly prompting the other.

\bibliographystyle{naturemag_davide}
{
\subsection*{References}\vspace{-0.6cm}
{\rm \bibliography{hierreview}}
}

\noindent

\subsection*{Acknowledgements} D.G. is supported by European Union's H2020  ERC Starting Grant No. 945155--GWmining, Leverhulme Trust Grant No. RPG-2019-350, and Royal Society Grant No. RGS-R2-202004.
M.F. is supported by NASA through the NASA Hubble Fellowship grant HST-HF2-51455.001-A awarded by the Space Telescope Science Institute.
Computational work was performed on the University of Birmingham BlueBEAR cluster, the Athena cluster at HPC Midlands+ funded by EPSRC Grant No. EP/P020232/1, and the Maryland Advanced Research Computing Center (MARCC).

\subsection*{Author Contributions} This paper was written jointly by the two authors. D.G. prepared the figures. M.F. extracted distributions from LIGO/Virgo public data.

Correspondence and requests for materials should be addressed to D.G. The authors declare no competing interests.

\end{document}